# Speculations on dark matter as a luminiferous medium

**Akinbo Ojo**
Standard Science Centre, P.O. Box 3501, Surulere, Lagos, Nigeria.
Email: taojo@hotmail.com

**Abstract.** Assuming the features commonly ascribed to dark matter, we discuss our glimpse of the possibility that dark matter as a light transmitting medium played a role that accounts for the discordant inferences about the dynamical behaviour of light, in which some optical experiments cannot detect the earth's motion in space, while others definitely can. In a converse way, the mentioned laboratory and astronomical observations provide an accessible experimental proof for the existence of a transparent, non-baryonic earth-bound medium which does not occupy the whole of universal space, dark matter being a possible candidate medium.

**Key words**: dark matter; light propagation; velocity of light.

PACS Classification: 95.35.+d

## 1. Introduction

Assuming gravity is the only relevant force and that the known gravitational laws do not need any significant modification on astronomical scales, the existence of dark matter in our galaxy in addition to the more easily perceived visible matter has become logical [1]. Among others, evidence for this inference includes the observed motion of outlying stars which shows that unless something extra is exerting a pull on them, galactic structure would not be intact as observed stellar speeds exceed the escape velocity based on the visible matter present. The amount of focusing that happens to light traversing a galactic cluster before reaching us also points to the existence of dark matter, as this is more than what can be accounted for solely by the material that can be seen directly.

Much current effort is concerned with identifying what dark matter is made of. That is, whether it is made of the ordinary atoms we are familiar with (i.e. baryonic matter), which are then constituted into brown dwarfs, white dwarfs and neutron stars or whether the matter is more exotic and non-baryonic. A lot of research effort is also spent in knowing whether the total amount is enough to make the cosmological parameter, $\Omega$ equal to unity.

While not wishing to distract those engrossed in these efforts, we wish to open another front for research by other speculators, namely what role dark matter would have to play as a light transmitting medium (i.e. as a luminiferous medium).

To do this, we recap in section 2 the common features ascribed to dark matter from gravitational theory and astronomical observations. We elaborate a bit more on an extra seventh feature concerning its distribution. In section 3 we briefly highlight the major experimental conclusions concerning the dynamical behaviour of light before discussing in section 4 whether any roles played by dark matter as a transparent medium can account for the seemingly discordant experimental findings about light's dynamical behaviour. Here, the features any potential luminiferous candidate must have are indicated, along with a mention of the difficulties faced by space vacuum and ether as potential candidates. Concluding remarks are made in section 5. This paper is unavoidably speculative, as also is the current state of the dark matter subject itself. Nevertheless, there are enough experimental grounds to support further research in the direction suggested.

**2. Features of non-baryonic dark matter**

The preponderant proportion of what constitutes dark matter is now known to be non-baryonic [2,3] and all subsequent reference to dark matter in this paper is to this dominant variety. Most of the features and their basis can be found in scientific literature, e.g. [1-3]. The following which are sufficient to discuss our topic are selected:

(i) Dark matter is largely inert as far as nuclear reactions are concerned, i.e. it is not composed of atoms, protons, electrons, etc. This feature is inferred from observed deuterium abundance and model calculations of primordial nucleosynthesis in the big bang.

(ii) Dark matter emits no light or other radiation.

(iii) Dark matter is transparent to transmitted light. That is, even though not made of atoms nor is it space vacuum, dark matter particles can transmit light waves.

(iv) Light is not significantly absorbed or scattered by dark matter. Thus, it has no colour.

(v) Dark matter can interact gravitationally with visible matter. It therefore has mass.

(vi) Observationally, the greater proportion, possibly over 90% of all the matter present within galaxies is dark matter.

(vii) The seventh feature of interest concerns how the dark matter within a galaxy is distributed. Detailed modelling of this distribution is already being researched, e.g. [4]. Our perspective here is provisional, so we discuss in a bit more detail. Is the dark matter smoothly distributed or does it exist in clumped forms within the galaxy?

Although some residual baryonic matter may be thinly spread in intervening spaces, the baryons within a galaxy are certainly not uniformly spread but clumped by gravity into planet and stellar sized celestial objects. Taking a cue from how baryonic matter is distributed, we can empirically generalize that rather than being uniformly spread, gravity does not discriminate between dark and visible matter but clumps all matter making up a galaxy into planet, stellar or larger sized objects with intervening space vacuum between them through which the objects can interact gravitationally.

According to models of galactic evolution, dark matter certainly existed in our galaxy alongside baryonic matter when the formation of our solar system started off within it. Three possible types of planet and stellar sized matter clumps can therefore be envisaged. A first type, which may consist almost purely of the dominant dark matter, in which case their presence can only be betrayed as unseen causes of perturbation or precession of planetary and stellar orbits. Then, a second variety can be conjectured which will be a mixture of dark and baryonic matter. And lastly, a third possible type consisting principally of the minority baryonic matter. Our earth consists of baryonic matter, so our planet could either belong to the mixed variety or the third type of clumped matter. On the basis of probability however, if 90% of our galaxy consists of non-baryonic matter we speculate that our earth would belong to the mixed variety.

Baryonic structure provides the guide that when attraction forces other than gravitational such as the strong and electromagnetic forces are present, additional clumping into hadrons (protons and neutrons), atomic nuclei, atoms and molecules take place within a gravitationally bound system. Taking a cue from this, we can logically infer that within those planet and stellar sized clumps composed of a significant mix of dark and baryonic matter, further clumping of the baryonic content will take place since the baryons therein can further interact with each other by means of attraction forces other than gravitational. The further aggregation of the baryonic content within such mixed types of matter clumps should therefore eventually result in a planet or stellar sized clump having a baryonic core and surrounded by a halo of non-baryonic dark matter.

Even if completely baryonic initially, it is not likely that throughout the period our earth has been evolving, it has been specially exempted from interacting with what now constitutes a very large percentage of the matter present in our galaxy. We therefore propose the earth as having its own share of dark matter interacting gravitationally with it and forming a halo around it, the baryonic earth being at its core.

Recent research papers on earth-bound dark matter [5,6] that have been brought to our attention are corroborative. In some of these a baryonic earth secondarily captures dark matter gravitationally as it wanders within the Milky Way galaxy. Using different methods, estimates of $\sim 10^{-8}$ [5] to $10^{-9}$ [6] of the earth's mass have even been given as the possible mass of this earth-bound dark matter halo.

**3. Inferences from experiments about the dynamical behaviour of light**
How light behaves dynamically in various transparent media though a well studied field has given rise to discordant inferences from the experimental facts obtained. Here, we mention some of the facts illustrative of the dilemma which physicists have confronted since early in the last century on how to correctly interpret light's dynamical behaviour. The experiments mentioned are not exhaustive and there are modern versions. A majority of these experiments were with a view to verify the existence of the hypothetical ether medium and discussions of them can be found in [7], excluding that of Sagnac which was published three years later [8].

Fizeau's experiment: The state of motion of a matter medium, e.g. water, relative to an observer has effect on the dynamics of light. That is, the velocity of light travelling in a matter medium is hastened or delayed, if the medium is moving towards or away from an observer respectively. As a noteworthy corollary to this finding, for a matter medium that is stationary relative to the observer even if the medium is in motion in some other reference frame, the velocity of light in that medium is constant and is not hastened or delayed in arriving at the observer.

Michelson-Morley type experiments: The velocity of terrestrial light travelling to a stationary earth-based observer is unaffected by the observer's earthly motion through space. That is, the arrival of incoming terrestrial light cannot be hastened or delayed by the stationary observer's earthly motion towards or away from the incoming light. Bluntly stated, *optical experiments cannot be used to detect the earth's motion in space*.

The Michelson-Morley experiment was designed to detect the earth's orbital motion about the sun. However, given as we now know that the sun and the Milky Way are themselves in motion, the earth's velocity in space is actually more than ten times the 30 kms$^{-1}$ that was modestly being looked for. This implies that the results obtained in the Michelson-Morley type experiments are ten times even further indicative of nullity. This to an extent discredits arguments that perhaps the small values obtained in the experiments are significant enough to imply a non-nullity. Modern Michelson-Morley type experiments have also been conducted in terrestrial vacuum and the null results obtained are very important because they strongly suggest that any medium that can possibly remain to transmit light and be stationary to the earth-based observer in order to satisfy the corollary to Fizeau's finding must be non-baryonic in nature.

Sagnac experiment: The velocity and thus arrival time of terrestrial light travelling to a non-stationary earth-based observer can be hastened or delayed, if the observer's rotational motion is towards or away from the incoming light respectively. Modified Sagnac experiments have been carried out which seem to indicate that the Sagnac effect not only applies to rotational motion but can also be found with uniform motion not stationary to the earth [9]. Thus, *optical experiments can at least detect the observer's motion in space.*

Bradley's aberration: The light from a zenith star shows aberration due to the observer's motion. This finding however admits alternative interpretations that suit opposing views. For example, it has been used as support by advocates of light whose velocity can not be relative and which requires no medium for propagation and proponents of a medium which can make light velocity relative to that of the observer also claim aberration supports their view. Some of the arguments can be found in [7].

Apart from the Bradley aberration there are more recent astronomical observations that are relevant to a discussion of light's dynamical behaviour. We mention two.

Cosmic microwave background (CMB) radiation anisotropy [10]: A red-blue anisotropy is seen with this radiation which otherwise demonstrates a high degree of isotropy. This observed anisotropy has been suggested to be due to the earth's motion relative to the incoming light implying that this *optical observation can be used to detect the earth's motion in space.*

Pulsar timing measurements [11]: Pulsars are recognized as very accurate clocks emitting light beams at very regular intervals. To maintain accuracy of pulsar timing recordings it is found necessary to correct for the earthly motion towards or away from the already emitted and incoming light beams because the earth-based observer's velocity hastens and delays the arrival of pulsar light. Thus, this *optical experiment can be used to detect the earth's motion in space.*

The hypothetical ether medium was supposed to occupy all universal space. Unfortunately the laboratory experiments mentioned above were not able to conclusively confirm the ether's existence. In particular, if the ether filled up all space and carried light waves, then the motion of a stationary earth-based observer moving through it should be capable of causing a hastening or delay in arrival of incoming light, an effect which was what was looked for in the Michelson-Morley experiment but was not detected.

Unwilling to give up on the idea of an ether, it was proposed that if the ether around the earth were bound to it then the null result of Michelson and Morley would be explainable, this being the corollary to Fizeau's experiment. This proposal, though logical however suffered from severe difficulties. Firstly, since the ether was supposed to fill up all space, for Bradley's aberration to be observable, the proposal required that the ether bound to the earth be incapable of communicating the disturbance of its own translational motion to the neighbouring portions of far off ether transmitting light from the zenith star but not moving along with the earth. This was found to be unworkable.

Secondly, the mechanism by which the ether would be bound to the earth and dragged along at its full velocity appeared impossible. The dragging effect was predicated on the Fresnel drag formula by which a drag coefficient given as $(1 - 1/n^2)$ was multiplied with the velocity of the medium relative to the observer to determine the fraction by which light was dragged along by the medium, where $n$ is the medium's refractive index. A partial dragging seemed to be ruled out by the null results of the Michelson-Morley type experiments which required that the medium drags light at the full velocity of the moving earth. Dragging light at this full velocity would however require the medium having special properties such as an unnaturally high, if not infinite refractive index so that the drag coefficient can almost equal one. For the case of a non-refractive medium, i.e. one with $n = 1$, the drag coefficient will be zero and there would even be no dragging at all. In such a case, incoming light would arrive earlier or later depending completely on the observer's velocity towards it and the Michelson-Morley results would not be null.

The moral of these difficulties was that if there be any medium, which would actually seem to be required to reconcile the null Michelson-Morley findings with others such as that of CMB anisotropy and pulsar light observation, the relevant medium must firstly not fill up all space so that Bradley's aberration is observable. Secondly, Fresnel dragging cannot be the mechanism by which the medium would accompany the moving earth, this requiring an unnaturally high refractive index for full dragging. Thirdly, if any medium is required it cannot be baryonic in nature since when all baryonic matter is taken out of consideration and the Michelson-Morley experiment performed in terrestrial vacuum, the results still remain null.

Given these difficulties and the lack of unanimity on how light behaves dynamically, Einstein thought it expedient to do away with the ether concept and rely on the experimental facts of the Michelson-Morley experiment that the velocity of light is constant and independent of the state of motion of the observer, i.e. arrival times of incoming light depended only on the distance at emission but is not dependent on whether the observer was stationary, moving towards or away from the incoming light after it has been emitted. The debate and arguments have however not abated till today. This is not just because of some fanatical attachment to the ether, but because it was felt that the Michelson-Morley experiment should not be singled out and other experiments showing that optical experiments can be used to detect the earth's motion wished away.

Why then is light playing such tricks on us? Reproducible laboratory and astronomical observations mentioned above demonstrate the two possibilities of light always arriving at the observer with a constant resultant velocity which cannot be added to or subtracted from no matter the observer's own velocity and of light velocity being increased or reduced by the observer's own velocity towards or away from the incoming light. Can these discordant experimental facts be reconciled by some additional understanding? Being a hitherto unknown but significant participant in our universe at the time of the experiments, we next wish to speculate on the possibility that dark matter may have played some role in the seemingly discordant experimental findings.

**4. Dark matter, space and ether as luminiferous media**
The motion of outlying stars from which dark matter is inferred enables us to safely say that all space is not filled with dark matter. Were dark matter uniformly present without galaxies as it is within, the observed motion of the outlying stars and the light bending of the images of distant galaxies by intervening ones would not be the way it is. We can deduce then that light can travel through space devoid of any form of matter, whether visible or dark. Light can also travel through matter, if transparent.

From the plausibility that the earth has its own dark matter halo, whenever we claim to measure the velocity of light *in vacuo* within our terrestrial laboratories, we must humbly entertain the possibility that what we may actually be measuring is the velocity of light in the earth's dark matter halo. Taking this possibility as a reality, we can propose that: the velocity of light in terrestrial vacuum is the velocity of light in the earth's dark matter halo.

Using this possibility as a basis, we proceed to examine its implication on the assumptions underlying the experiments mentioned in section 3, hoping thereby to glimpse a reason why the conclusions should appear discordant.

The assumption of Michelson and Morley was that they were experimenting in a space which was of the same nature as all universal space. Since dark matter was still an unknown possible participant at that time this assumption is understandable. Fizeau's experiment on the other hand had previously shown that the presence and state of motion of a matter medium relative to the observer has an effect on how light behaves dynamically.

Now, if the experiment of Michelson and Morley contrary to their innocent assumption was an experiment carried out in a transparent matter medium, the earth's dark matter halo, and if this halo were at rest relative to the stationary earth-based observer, being gravitationally bound, then their experimental conclusions become understandably null and equivalent to the corollary of Fizeau's experiment noted above.

It also becomes understandable why light coming from beyond the earth's dark matter halo, such as CMB radiation and pulsar light, can be hastened or delayed due to the motion of the stationary earth-based observer towards or away from their incoming light, since these traverse an extra-terrestrial space medium which is not stationary to the earth-based observer, not being gravitationally bound or capable of being dragged along if its refractive index equals one.

For the non-stationary earth-based observer moving relative to the earth's dark matter halo, terrestrial light can be hastened or delayed by the observer's motion within the halo as medium and observer are not stationary relative to each other, which concurs with findings in the Sagnac and modified Sagnac experiments.

We are informed by Galileo [12] that if dropping water (representing light), the receiving bowl (representing the observer) are all within the same ship (representing a matter medium), the dropping of water into the bowl will be unaffected by the ship's uniform motion. Using this analogy to discuss Bradley's aberration, if the dropping water at least in an initial part of its journey to the bowl is not from within the ship, the dropping water must suffer aberration analogous to that observed by Bradley for light coming from a zenith star. If however the ship's cabin extends to the zenith star, no aberration will be observed. Again, if there is only a moving bowl but no ship, aberration of the dropping water will still be observed. Aberration can therefore not be used to confirm the presence or absence of a ship, its observation only confirms the bowl's movement and that the dropping water is not from within the ship's cabin, should there be a ship. This interpretation has been borne out experimentally [13], where it has been demonstrated that the presence or absence of a matter medium does not distort the observance of aberration, in as much as the matter medium, if present does not extend to the zenith star being observed.

Like the luminiferous ether, dark matter is transparent and non-baryonic. However, unlike the ether, dark matter does not fill up all space and this being so, it does not suffer much of the difficulty of how to avoid communicating its translational motion to any possible nearby medium transmitting light from a zenith star and so Bradley's aberration remains observable.

Again unlike the ether, dark matter being in the same frame of motion as the moving earth is not as a result of being dragged along according to Fresnel's formula but because it forms a single gravitational object with the earth. Dark matter would therefore not require any special optical properties for it to accompany the baryonic earth at its full velocity.

Finally, since removing baryonic matter from the light path in the Michelson-Morley type experiments may not necessarily imply removal of non-baryonic matter, dark matter can remain to provide a medium that can make the results stay null. Taken together these characteristics of dark matter make it a very strong candidate medium for reconciling the seemingly discordant inferences about light's dynamical behaviour.

Any alternative candidate medium must similarly possess the capability of being earth-bound without the need for Fresnel dragging which would require an unnatural refractive index to accompany the earth at its full velocity. Secondly, it must not fill the whole of space so that Bradley's aberration is observable. Thirdly, it must be non-baryonic in nature so that the Michelson-Morley results remain null in terrestrial vacuum.

Space vacuum as a luminiferous medium cannot by itself without some additional understanding provide a medium that accounts for the discordant experimental findings. In particular, it cannot accompany the moving earth to satisfy the corollary to Fizeau's finding above because firstly, it cannot be gravitationally bound and secondly, it cannot be dragged along if its refractive index equals one, yet the discordant experiments require either of the two mechanisms for their reconciliation.

The ether as a candidate can satisfy the condition of being non-baryonic. However, if it can only accompany the earth by Fresnel dragging this requires a special refractive index to fully drag the light. Alternatively, if it has mass and can be gravitationally bound, its occupying all of space would make it a single gravitational object with the earth. This would make Bradley's aberration unobservable since all the ether and the earth will be bound in same frame of motion. Since there can be many more earth-like bodies in space, this would also make our earth special. Resolving the ether's difficulties would probably require the ether to be of two variants, one capable of being bound and interacting gravitationally and the other gravitationally inert. In such a scenario, the former variant would still be equivalent to a dark matter.

**5. Concluding remarks**
On the premise that dark matter is present within our galaxy and that like all matter it is not smoothly distributed but clumped by gravity, added to the further belief that our earth is not exempted from interacting gravitationally with what may constitute the major proportion of the matter in our galaxy, we speculate the possibility here that dark matter was an unknown participant in the early terrestrial laboratory experiments concerning the behaviour of light.

Since light is influenced by the dynamics of the matter medium through which it travels as demonstrated by Fizeau's experiment, light dynamics must then be influencible by the presence of a dark matter medium.

The participation of a dark matter medium forming a halo around the earth can reconcile the seemingly discordant experimental conclusions of Fizeau, Michelson-Morley, Sagnac, pulsar timing measurements and CMB anisotropy observations, whereby the earth's motion is not discernible at all in some optical experiments while it is definitely detectable in others.

In a converse perspective, the above experimental findings about light provide an accessible experimental proof for the existence of a transparent, earth-bound medium which does not fill the whole of universal space and which does not have a baryonic nature.

If the speculations here are correct, they must have some bearing on the Special theory of relativity. Whether dark matter would also have any bearing on some phenomena predicted by General relativity such as the precession of orbits, the bending and change in the dynamics of light grazing the surface of a celestial body or passing through a galactic cluster are future study areas.

**Acknowledgement**
We thank J. Dunning-Davies for drawing our attention to a recently published paper on earth-bound dark matter [6] and for his caveat that the dark matter idea holds only if gravity alone is the relevant consideration.